\documentclass[12pt]{article}
\usepackage{cite}
\usepackage{graphicx}
\usepackage{geometry}                
\pdfoutput=1

\newcommand{\beq}{\begin{equation}}
\newcommand{\eeq}{\end{equation}}
\newcommand{\ie}{{\it i.e.}}
\newcommand{\eg}{{\it e.g.}}
\newcommand{\eq}[1]{Eq.\,(\ref{#1})}

\newcommand{\eqsto}[2]{Eqs.~(\ref{#1}) to (\ref{#2})}

\renewcommand{\epsilon}{\varepsilon}

\newcommand{\gsim}{\stackrel{>}{_\sim}}

\newcommand{\clean}{^{\rm clean}}
\newcommand{\sat}{^{\rm sat}}
\newcommand{\clogged}{^{\rm clog}}
\newcommand{\binding}{_{\rm bind}}
\newcommand{\colision}{_{\rm collide}}
\newcommand{\trapping}{_{\rm trap}}
\newcommand{\dd}{\mbox{d}}
\newcommand{\ene}{\mbox{$n$}}

\newcommand{\ze}{\mbox{$z_e$}}
\newcommand{\rcero}{\mbox{$r_0$}}
\newcommand{\runo}{\mbox{$r_1$}}
\newcommand{\re}{\mbox{$r_e$}}
\newcommand{\reclean}{\mbox{$r_e\clean$}}
\newcommand{\resat}{\mbox{$r_e\sat$}}

\newcommand{\rhocero}{\mbox{$\rho_0$}}
\newcommand{\rhouno}{\mbox{$\rho_1$}}
\newcommand{\rhoe}{\mbox{$\rho_e$}}
\newcommand{\rhoeclean}{\mbox{$\rho_e\clean$}}
\newcommand{\rhoesat}{\mbox{$\rho_e\sat$}}
\newcommand{\deltax}{\mbox{$\delta x$}}
\newcommand{\omegacolision}{\mbox{$\Omega\colision$}}
\newcommand{\omegabinding}{\mbox{$\Omega\binding$}}
\newcommand{\omegabindingclean}{\mbox{$\Omega\binding\clean$}}
\newcommand{\omegabindingsat}{\mbox{$\Omega\binding\sat$}}
\newcommand{\omegatrap}{\mbox{$\Omega\trapping$}}
\newcommand{\omegacero}{\mbox{$\Omega_0$}}
\newcommand{\omegauno}{\mbox{$\Omega_1$}}
\newcommand{\zcero}{\mbox{$z_0$}}
\newcommand{\zclean}{\mbox{$z_0$}}
\newcommand{\zsat}{\mbox{$z_e\sat$}}
\newcommand{\zuno}{\mbox{$z_1$}}
\newcommand{\tred}{\mbox{$t'$}}
\newcommand{\tredhalfsat}{\mbox{$t'_{1/2}$}}
\newcommand{\enered}{\mbox{$\ene'$}}
\newcommand{\eneclean}{\mbox{$\ene\clean$}}
\newcommand{\enesat}{\mbox{$\ene\sat$}}
\newcommand{\eneclogged}{\mbox{$\ene\clogged$}}
\newcommand{\flujo}{\mbox{$\Phi$}}
\newcommand{\flujoimpurezas}{\mbox{$\flujo_{\rm imp}$}}

\newcommand{\Cimpurezas}{\mbox{$C_{\rm imp}$}}
\newcommand{\Aredlin}{\mbox{$A'_{\rm lin}$}}
\newcommand{\Aredlog}{\mbox{$A'_{\rm log}$}}

\begin{document}

\newcommand{\titulo}{An effective-charge model for the\\ trapping of  impurities of fluids\\ in channels with nanostructured walls}
\newcommand{\autor}{M.V.~Ramallo}
\newcommand{\direccion}{LBTS, Departamento de F\'{\i}sica da Materia
Condensada,\\ Universidade de Santiago de Compostela, ES-15782 Santiago de Compostela, Spain}

\mbox{}\vspace{2cm}  \mbox{}

\begin{center}
  \Large\bf
\titulo\\  \end{center}\mbox{}\vspace{-1cm}\\ 

\begin{center}\normalsize\autor\end{center} 

\begin{center}\normalsize\it\direccion\end{center}


\mbox{}\vskip0.5cm{\bf Abstract. }
We present model equations for the trapping and accumulation of particles in a short cylindrical channel with nanostructured inner walls when  a fluid passes through, carrying a moderate load of impurities.  The basic ingredient of the model is the introduction of a phenomenological ``effective-charge density'' of the walls, related to the electrical charges exposed in the nanotexture, and which is gradually reduced as the flow runs through the channel and the trapped impurities cover the internal walls. By solving the proposed equations, three regimes are predicted for  the  channel:  a linear or  clean-filter regime, a logarithmic or half-saturation regime, and the saturation limit. It is proposed that experimentally testing these regimes may help to understand the enhanced trapping capability observed in many diverse nanotextured channel structures.\vspace{1cm}

\mbox{}\hfill{\footnotesize {\tt ramallo@cond-mat.eu}}
\thispagestyle{empty}

\newpage
\setlength{\baselineskip}{18pt}


\newpage
\setlength{\baselineskip}{18pt}


\section{Background introduction}
Recently,\cite{nanotubos-nature,graphene,review-nanofilters-2011,review-nanofilters-2008,virus-a,virus-b,argonide-1,argonide-2}  researchers from both academia and industry have experimentally demonstrated  that a variety of nanostructures and nanotextured media (see some examples later in this introduction) can  efficaciously trap nanoimpurities carried by fluids when a flow is induced by external  hydrostatic pressure. These findings are not only   scientifically interesting, but  also  promising  for the socially and economically important application of  purification of drinking water,\cite{review-nanofilters-2008,reporteconomico} and of other liquids\cite{argonide-2}. When compared to conventional porous filters, the new media have the important advantages of retaining impurities of sizes typically in the  tens of nanometers and, at the same time, presenting a resistance to hydrodynamic flow orders of magnitude smaller than what conventional  models would predict for  channels of diameters as small as the  particles being trapped.  

Roughly, we can  divide the structures presenting such enhanced impurity trapping capability into two groups: {\it a)}~Those formed by nanometric-diameter channels through which the fluid  flows.\cite{nanotubos-nature,graphene,review-nanofilters-2011,review-nanofilters-2008} A well known  example is the  nanotube arrays grown and experimentally tested by Srivastava and coworkers.\cite{nanotubos-nature}  Other specially interesting example  are graphene membranes, although by now  they have been probed only through molecular dynamics simulations.\cite{graphene} In any nanometric-diameter channel, simple size exclusion will play a major role in the retention of nanoimpurities. Still, these structures   exhibit remarkable trapping capability also for some ions significantly smaller than the channels' diameter.\cite{nanotubos-nature,graphene}  The resistance to flow is observed to be well lower than what conventional models predict for these diameters, a phenomena often attributed to water-nanostructure interactions (see, \eg, \cite{nanotubos-nature}) though not yet fully understood at the quantitative calculation level. {\it b)}~The second group corresponds to nanostructures embedded in larger  structures, resulting in filters composed by channels with micrometric diameters and inner walls coated with nanoparticles. Examples are conventional microfilters   coated with Y$_2$O$_3$\cite{virus-a}, ZrO$_2$\cite{virus-b},  or Al$_2$O$_3$~\cite{argonide-1,argonide-2} nanopowders (further examples can be found in the reviews \cite{review-nanofilters-2011,review-nanofilters-2008,reporteconomico}). These structures have been observed by their growers to have a surprisingly good filtration performance for nanometric impurities, as small as $\sim10$~nm, in spite of the relatively large diameter of the channels (note that in a channel of diameter 1~$\mu$m only about 0.04\% of the fluid will transit closer than 10~nm from the walls).\cite{virus-a,virus-b,argonide-1,argonide-2,review-nanofilters-2011,review-nanofilters-2008,reporteconomico} Their hydrodynamic resistance is quite low, similar to the one of conventional micrometric filters. The filtration mechanism of these structures is unknown as yet. Nonetheless, the initial trapping capability was observed to depend on pH,\cite{virus-a,virus-b,argonide-1} and thus electrostatic and polar attraction may be suspected to play a  role.

\begin{figure}[b!]
\begin{center}\mbox{}\vspace{1.5cm}\\
\includegraphics[width=\textwidth]{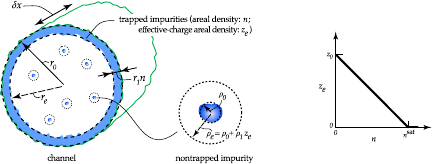}\end{center}
\caption{Representation of a nanostructured  channel  filter as modelled in the present Letter. The radiuses \rcero\ and \rhocero\ correspond to the  average dimensions of the bare channel and  impurities. The effective radiuses \re\ and \rhoe\ vary as trapped impurities cover the inner wall, via their dependences on, respectively, the areal density \ene\ of trapped impurities and on the areal density \ze\ of effective charge of the inner wall. This \ze\ reflects that exposed charges in a nanostructured surface  attract  the impurities in the fluid and also constitute binding anchors for those impurities. It is expected to diminish as impurities cover the surface, for which  we assume the simple $\ze(\ene)$ dependence  plotted at the right (although it can be  easily generalized to more complex functionalities, see main text).}\label{fig.1}
\end{figure}

The purpose of this Letter is to propose a semi-macroscopic model  for the nanoimpurity trapping capability of   short cylindrical-like  channels with nanostructured inner walls of the type composing filters of category {\it  b} in the previous paragraph. The equations do not assume any detailed chemical compound for the nanotexture, in correspondence with the ample variety of nanostructures producing the  enhanced impurity trapping capability. The model produces agreement with the  trapping performances quantitative reported by experimentalists in various systems. These performances correspond always to the initial filtration behaviour, \ie, they were always measured in the clean-channel limit. In order to further explore the model, we also calculate the time-evolution of the  trapping performance and impurity accumulation as flow continues and the channels become dirtier. This allows us to  propose future measurements that may   test these ideas more thoroughly.  We believe that some aspects of the model  could  be also useful to explain in part the trapping of the smaller ions in  the nanodiameter channels of category {\it a}. But  its full applicability to that case is limited by our use of classical  dynamics for the carrying fluid, and hence we do not focus here on that category (also, for  these nanodiameter channels, in which the number of fluid atoms is manageably small,  molecular dynamics simulations  as those in \cite{graphene} could be a more reliable, albeit not general, approach).

Our theoretical proposal explores a simplified (but realistic) view, in which the improved filtration capability is primarily due to the fact that  the nanotexturing  exposes electrical charges in the walls which induce both electrostatic and van der Waals attraction over the impurities in the fluid. This nanostructuring also provides chemical anchors for the binding of those impurities once they collide with the channel walls. Correspondingly, our basic ingredients will be the introduction of an effective-charge density, \ze, of the inner walls of the channels, and writing down  as a function of  \ze\  the probabilities for impurity collision and binding in the channel walls, \omegacolision\ and \omegabinding. As it could be expected, \ze\ will vary with the  areal density \ene\ of impurities trapped in the inner walls of the channel.  We then obtain within the model the evolution of \ene\ with time $t$, as  (for the fully explicit form of this equation, see   \eq{diferencialred} later in this Letter):
\beq
\frac{\dd\ene}{\dd t}\;\propto\;\flujoimpurezas(\ene)\;\;\omegacolision(\,\ene,\ze(\ene)\,)\;\;\omegabinding(\,\ze(\ene)\,),
\label{eq-intro}
\eeq
 where \flujoimpurezas\ is  the impurity flow from the incoming fluid.  This equation is nonlinear. By  solving it both numerically and through analytical approximations, we  predict three different regimes on the $\ene(t)$ evolution:  a linear or  clean-filter regime, a logarithmic or half-saturation regime, and the saturation limit (see Fig.~2a). 
 
\mbox{}
 
\section{Obtainment of an equation for the areal density of trapped impurities in a channel with nanostructured walls}

\subsection{Initial modelling and notations} 

Our starting point, and most of our  basic notations, are illustrated in Fig.~1. We consider a channel with nanostructured inner walls, its nominal shape being cylindrical-like with average  radius \rcero\ and length \deltax. Inside this channel a fluid flows due to externally-applied hydrostatic pressure, carrying a load of impurities. Some of those impurities will become trapped by the inner wall of the channel, then reducing its effective radius to a value $\re=\rcero-\runo\ene$, where \ene\ is the number areal density of trapped impurities in the inner wall and \runo\ is a constant of the order of the average radius of impurities, \rhocero.  All through this Letter, by ``areal density'' we refer to quantities normalized using the nominal area of the inner wall, $2\pi\rcero\deltax$  (and not  the cross-section of the channel). Also, for simplicity we consider all impurities equal among them (subsequent generalization to multiple chemical species should be easy) and the length \deltax\ short enough as to  take \ene\ and \rcero\ constant through the channel's axis coordinate (generalization to longer lengths, and also to some different shapes, should also be easy, by integration over \deltax). The impurity concentration in the fluid is considered to be moderate enough as to not significantly affect its viscosity, and as for the impurities in the fluid to be noninteracting with each other (specially when colliding with the inner wall).

\subsection{Effective-charge density of the inner wall, \ze}  

We now introduce the important concept of a phenomenological ``effective charge'' of the inner wall of the channel. We quantify this effective charge via its areal density \ze\ and, as already commented on in the introduction, it reflects the fact that the nanostructured  walls expose charges that induce both electrostatic and van der Waals attraction over the components of the impurities in the fluid. Indeed,  \ze\ will depend on the areal density of already trapped impurities \ene\ (which will screen out the wall) and also on the chemistry specifics of the wall and impurities.  Let us focus  on the mutual interplays between \ene\ and \ze\ and in obtaining an equation for their evolution with time as flow passes through the channel. In particular, the interdependence $\ze(\ene)$ may be naturally expected to be continuously decreasing when \ene\ increases, to take a finite value \zclean\ at $\ene=0$ (clean filter), and to saturate to zero when \ene\ reaches some critical value \enesat\ at which all active centers of the wall become well covered by impurities. We thus postulate the simplest $\ze(\ene)$ dependence fulfilling such conditions (plotted in Fig.~1):
\begin{eqnarray}
\ze(\ene)&=&\left\{
\begin{array}{lr}
\zclean\left(1-\frac{\ene}{\enesat}\right),& {\rm if}\;\ene<\enesat\\
0,& {\rm if}\;\ene\geq\enesat
\end{array}
\right.\nonumber\\
&=&\zclean\left(1-\Big\Vert\frac{\ene}{\enesat}\Big\Vert\right),
\label{zn}
\end{eqnarray} 
where the notation $\Vert...\Vert$ stands for $\min\{1,...\}$. Obviously other sensible choices for $\ze(\ene)$ are possible such as, \eg, $\ze(\ene)=$$\zclean(1-\Vert\ene/\enesat\Vert)^{\zuno}$, with \zuno\ a positive coefficient that probably covers at a good approximation most actual possibilities depending on its value, as it corresponds  to $\ze(\ene)$ downwards curvature if $0<\zuno<1$,  to no curvature for $\zuno=1$ (\ie, \eq{zn}), and to upwards curvature for $\zuno>1$. For simplicity, we shall consider in this Letter only the case $\zuno=1$.
 
\subsection{Impurity trapping probabilities as a function of \ze\ and \ene}  The role played by \ze\ in our model will be in fact two-fold. First, it affects  how large is the distance within which if the impurity approximates the inner wall the latter attracts the former so much as to consider it as a  collision. This attraction distance may be  seen as an effective radius, \rhoe, of the impurity (see Fig.~1), so that if the distance from the center of the impurity to the center of the channel is larger than $\re-\rhoe$ the impurity will actually touch the wall (dressed with already trapped impurities). For concreteness, we will assume that the dependence of \rhoe\ on \ze\ can be linearized, in the relevant range of values, as $\rhoe=\rhocero+\rhouno\ze$, where \rhouno\ is a constant. (In fact, an exact linear relation can be easily obtained on the grounds of simple energy balances using an electrostatic interaction and with the thermal energy playing the role of an escape kinetic energy, suggesting then also that \rhouno\ will be roughly inversely proportional to temperature.) Using this \rhoe\ we may now obtain the probability \omegacolision\ that, at a given instant, a given flowing impurity will collide with the impurity-dressed wall. For that, we assume that the concentration of impurities is constant in all the transversal section of the fluid and that the fluid velocity profile is given by the Poiseuille law,\cite{Poiseuille} $u(r)\propto\re^2-r^2$, where $u$ is the fluid velocity and $r$ the distance to the channel's axis (see Ref.~\cite{review-micro1} for an explicit discussion supporting that at least for channels of radius $\gsim10\,\rm{nm}$ the flows of water-like liquids driven by hydrostatic pressure are in fact in the Poiseuille regime). The probability \omegacolision\ is then given by the fraction of the fluid mass that passes through the outer ring given by $\re-\rhoe\leq r\leq\re$, \ie, $\omegacolision=\int_{\re-\rhoe}^{\re} u(r) r \dd r / \int_{0}^{\re} u(r) r \dd r$. The result of those integrations is: 
\begin{equation}
\omegacolision=
\Big[\Big(\Big\Vert\frac{\rhoe}{\re}\Big\Vert-1\Big)^2-1\Big]^2.
\end{equation} 
Note that, through the $\re(n)$ and the $\rhoe(\ze)$ dependences, \omegacolision\ will vary as flow passes and the walls become dirtier. 

The second influence played by \ze\ in our model concerns the probability that an impurity gets actually bound to the inner wall of the channel once it actually is within a collision distance from that wall. The probability of such a process, per unit length, will be denoted here as \omegabinding. Obviously this probability will depend on the chemistry of  impurities and  active centers of the nanostructure, and also on the number of  active centers not yet saturated by existing bindings. The latter indicates that \omegabinding\ will grow with \ze, and in particular we may adopt  the natural first-order approximation $\omegabinding=\omegacero+\omegauno\ze$ (\omegacero\ reflects then a conventional  binding probability in a conventional non-nanostructured filter and $\omegacero\ll\omegauno\zcero$). Note that the probability \omegatrap\ that a given impurity in the fluid flow gets trapped in the walls during its transit trough the channel  is then given by $\omegatrap=\omegacolision\omegabinding\deltax$,  expressed above as a function of \ene\ and \ze.
 
\subsection{Equation for $\dd\ene/\dd t$} Let us now build, on the basis of the above relationships, an equation for the evolution of the areal density of trapped impurities, \ene, as a function of time when an impure fluid flows through the channel due to hydrostatic pressure. In fact, $\dd\ene/\dd t$ itself is an almost equally important quantity, as it gives the filtration capability of the channel. This derivative may evidently be expressed as $\dd\ene/\dd t=(\flujoimpurezas/2\pi\rcero\deltax)\omegatrap$, where \flujoimpurezas\ is the flow of impurities brought by the incoming fluid (in units of ${\rm s}^{-1}$; the factor $(2\pi\rcero\deltax)^{-1}$ is due to the areal density normalization in the definition of \ene). Gathering together the previous results in this Letter, this can be written now as:
\begin{equation}
\frac{\dd\ene}{\dd t}=\frac{\flujoimpurezas(\ene)}{2\pi\rcero}
\;\Big[\omegacero+\omegauno\ze(\ene)\Big]\;\Big[\Big(\Big\Vert\frac{\rhocero+\rhouno\ze(\ene)}{\rcero-\runo\ene}\Big\Vert-1\Big)^2-1\Big]^2,
\label{diferencial}
\end{equation} 
with $\ze(\ene)$  given by \eq{zn}. This \eq{diferencial} may already be integrated if we further assume $\flujoimpurezas(\ene)$ to be constant with \ene\ or, equivalently, with time. However, we believe that for the experiential applications of this equation the most interesting case will be the one of constant pressure difference $P$ between both ends of the channel. Then \flujoimpurezas\ will depend on \ene\ and, in particular,  as the channel gets dirty it may become clogged enough as to significantly  decrease  the flow of the fluid. For this case of constant $P$, the $\flujoimpurezas(\ene)$ dependence can be  obtained by employing again the Poiseuille law for fluid flows driven by hydrostatic pressure in cylindrical channels, that relates the channel's dimensions to its resistance to flow as $\flujo=\pi P\re^4/8\eta\deltax$, where $\flujo=\flujoimpurezas/\Cimpurezas$ is the fluid flow,  \Cimpurezas\ is the incoming number concentration of impurities, and $\eta$ the viscosity of the fluid. For the case of constant $P$ and \Cimpurezas\  we then get:
\begin{equation}
\flujoimpurezas(\ene)=\;
\frac{\pi\Cimpurezas P}{8\eta\deltax}\;\big(\rcero-\runo\ene\big)^4.
\label{flujo}
\end{equation} 
(Note that for $\ene=\eneclogged\equiv\rcero/\runo$  it is $\flujoimpurezas=0$ and $\re=0$, \ie, the channel becomes fully closed by impurities at that \ene-value).

While \eqsto{zn}{flujo}  form now a closed set of equations that can be already solved to determine $\ene(t)$, it will be very useful to first reexpress them more compactly, and in terms of a more recognizable set of physical quantities. In particular, and as it could have been expected, it turns out that the filtering behaviour  can be described in terms of its features at two specific instants of its evolution: Its clean point, defined by $\eneclean=0$, and its saturation point, defined by $\ene=\enesat$ (\ie, $\ze=\zsat=0$). In particular, \eqsto{zn}{flujo} can be rewritten as
\begin{equation}
\begin{array}{c}
\frac{\dd\enered}{\dd\tred}=(\enesat\reclean)^{-1}\;\Big[\reclean-\;\left(\reclean-\resat\right)\;\enered\Big]^4\;\times\vspace{0.5em}\\
\Big[\omegabindingsat+\,\big(\omegabindingclean-\omegabindingsat\big)\big(1-\Vert\enered\Vert\big)\Big]\;\times\vspace{0.5em}\\
\Bigg[\Big(\Big\Vert\frac{\rhoesat\;+\;\big(\rhoeclean-\rhoesat\big)\mbox{$(1-\Vert\enered\Vert)$}}{\reclean\;-\;\big(\reclean-\resat\big)\,\enered}\Big\Vert-1\Big)^2-1\Bigg]^2,
\end{array} 
\label{diferencialred}
\end{equation} 
\mbox{}\\
where $\enered=\ene/\enesat$ and $\tred=t/\,(16\eta\deltax/\Cimpurezas P)$ are the so-called reduced density of trapped impurities and reduced time, respectively. (In fact, already  a first result of our equations is that the quantities $\eta$, \deltax,  \Cimpurezas\ and $P$ only affect the impurity trapping behaviour in the time scale, as they enter \eq{diferencialred} only through the \tred-normalization.) The other parameters in \eq{diferencialred} are the effective radiuses of  channel and impurities, and the binding probabilities, in the clean and saturated points:
\begin{equation}
\left\{
\begin{array}{rcl}
\reclean&=&\rcero,\\
\rhoeclean&=&\rhocero+\rhouno\zcero,\\
\omegabindingclean&=&\omegacero+\omegauno\zcero,\\
&&\\
\resat&=&\rcero-\runo\enesat,\\
\rhoesat&=&\rhocero,\\
\omegabindingsat&=&\omegacero.
\end{array}
\right.
\end{equation} 
These 6 variables are thus equivalent to the set $\{\rcero$, \runo, \rhocero, $\rhouno\zclean$, \omegacero, $\omegauno\zclean\}$ and  together with \enesat\ they define the $\enered(\tred)$ behaviour through \eq{diferencialred}, as will be discussed in the next section. (Actually, in all rigour these 7 degrees of freedom could  be further reduced to 6 by rescaling them  to a single reference length, such as \rhocero, \ie, \rhoesat).
\begin{figure}[b!]
\begin{center}\mbox{}\vspace{1.5cm}\\
\includegraphics[width=\textwidth]{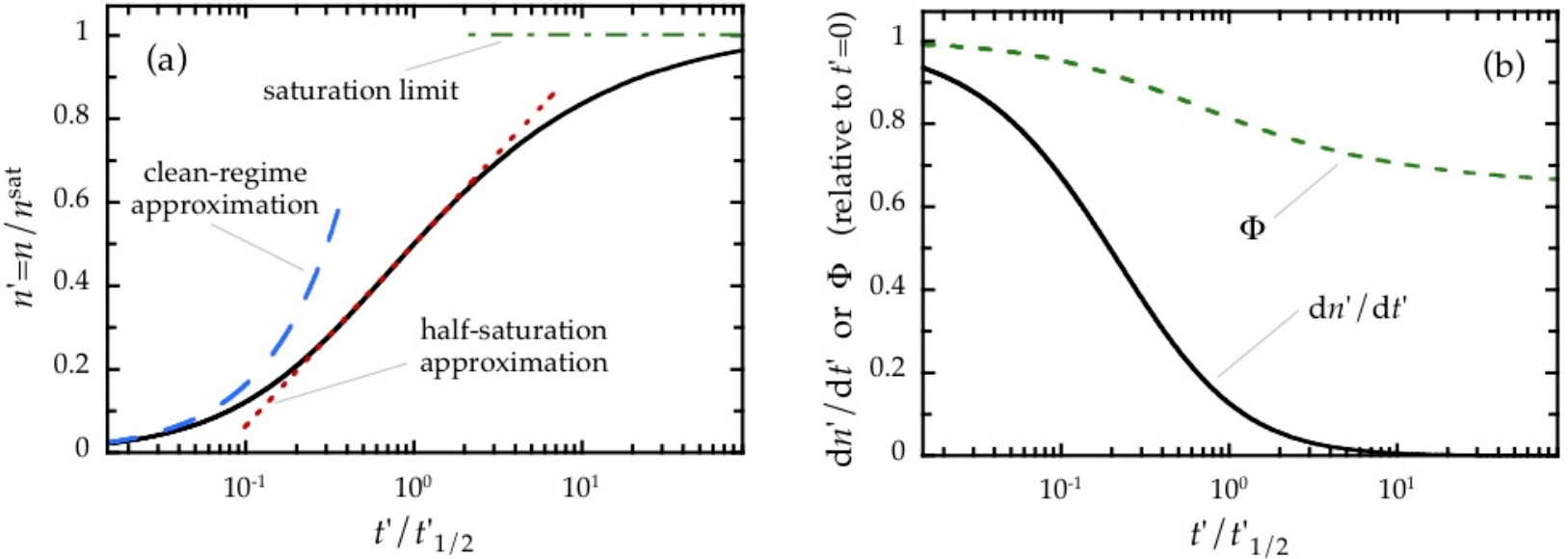}\end{center}
\caption{~(a):~Results, obtained by integrating  \eq{diferencialred}, for the areal density of trapped impurities in units of the saturation value, $\enered=\ene/\enesat$ (continuous line), as a function of time in units of the half-saturation value, $\tred/\tredhalfsat$ with \tredhalfsat\ defined by $\enered(\tredhalfsat)=0.5$. The parameter values used are (see main text for details): $\rhoesat=\rhocero=13$\AA,  $\rhoeclean=20\rhocero$, $\resat=450\rhocero$, $\reclean=500\rhocero$, $\enesat=5/\rhocero^2$, $\omegabindingsat=0$, $\omegabindingclean=0.5/10^3\rhocero$, and the initial value  was $\enered(0)=0$. We also show  the results obtained  from the clean-regime approximation, \eq{clean-1} (dashed line),  from the half-saturation approximation, \eq{sat-1} (dotted line), and for the saturation limit, $\ene=\enesat$ (dot-dashed line).  (b):~Filtration capability $\dd\enered/\dd\tred$ (continuous line) and fluid flow \flujo\ (dashed line),  normalized to their values at $t'=0$ and as a function of $\tred/\tredhalfsat$, obtained again from  \eq{diferencialred} and the same parameter values as in Fig.~2a. Note that the horizontal axes of these plots are logarithmic, so that the half-saturation regime is well longer than the clean regime. The hydrodynamic flow through the channel is reduced only about 30\% during its working lifetime.}
\label{fig.2}
\end{figure}

\mbox{}

\section{Discussion of the $\enered(\tred)$   obtained by integrating the model equation}

\subsection{Numerical integration}

We show in Fig.~2(a) an example of the $\enered(\tred)$ results obtained by integrating \eq{diferencialred} using some representative values for the  parameters involved (and always in the case of constant $P$ and \Cimpurezas, and starting from a clean initial state $\enered(0)=0$). The computation is lightweight and takes less than a minute in current personal computers. In particular, we have chosen for the impurity sizes $\rhoesat=\rhocero=13$\AA\ and $\rhoeclean=20\rhocero$; for the channel radius sizes $\reclean=500\rhocero$ and $\resat=450\rhocero$; for the saturation impurity density $\enesat=5/\rhocero^2$; and for the binding probabilities $\omegabindingclean=0.5/10^3\rhocero$ and $\omegabindingsat=0$ (so that we neglect conventional filtration mechanisms and focus on the effects of nanostructuring alone). We have chosen these values because they reproduce the nominal sizes and the initial filtration capability experimentally reported for channels coated with  Y$_2$O$_3$ nanopowders\cite{virus-a} and also for channels coated with  ZrO$_2$ nanopowders\cite{virus-b}. In both cases, channel arrays of length $\sim7$~mm have been show to retain MS2 viruses (of radius  $\sim13$~nm) with an initial logarithmic reduction value of 7-log when the channels are still clean. Note that the initial   logarithmic reduction value in a finite-length clean channel can be simply estimated in our model  through the accumulated non-trapping probability as $-(L/\rhocero)\log_{10}[1-\rhocero\omegabindingclean[(\rhoeclean/\reclean-1)^2-1]^2]$, where $L$ is the length of such channel. With the above choice of parameter values this estimate yields the measured\cite{virus-a,virus-b} 7-log impurity trapping. Obviously a more stringent determination of the parameter values for these example filters would need further measurements not only of the initial filtration performance but also of its detailed evolution with time.  Also obviously, other example channels may correspond to  different parameter values: for instance, to reproduce the 94\% retention rate of MS2 viruses reported in \cite{argonide-1} in channel arrays with  dimensions   similar to the ones in \cite{virus-a,virus-b} but with a 10wt\%-Al$_2$O$_3$ nanocoating, the choice $\omegabindingclean=0.8/10^4\rhocero$ would be more appropriate. 

As show in Fig.~2(b), we have also computed, using the same parameter values as for Fig.~2(a), the  evolution with time of both the fluid flow \flujo\ and the derivative $\dd\enered/\dd\tred$ that gives a measure of the filtration effectiveness at each instant. Note that \flujo\  experiences only a moderate reduction during the filtering working lifetime,  other of the important  features   observed by industry researchers\cite{argonide-1,argonide-2} and of paramount importance for applications.

\subsection{Clean, half-saturation and saturation regimes}
At least three regimes  can be easily noticed in Fig.~2(a): {\it}~The initial behaviour or clean regime, at which the  growth of $\enered(\tred)$ is approximately linear; {\it ii)}~an intermediate regime, henceforth called half-saturation regime, where the  growth of $\enered(\tred)$ is approximately logarithmic; and {\it ii)}~the saturation limit in which \ene\ approaches a value \enesat\ at a slow pace. In fact, in that figure we also show as dashed or dotted lines the simple analytic approximations for $\enered(\tred)$ that can be easily found in the first two of these regimes by introducing in \eq{diferencialred} the corresponding approximations  $\tred=0$ and, respectively,  $\enered=1/2$. This equation becomes in those approximations as follows:
\begin{equation}
\left\{
\begin{array}{ccl}
\enered(\tred)&\simeq&\Aredlin\tred\\
&&\\
\frac{\dd\enered}{\dd\tred}&\simeq&\Aredlin
\end{array}
\right.
\mbox{(linear or clean regime $\tred\simeq0$),}
\label{clean-1}
\end{equation} 
with
\begin{equation}
\Aredlin=\frac{\omegabindingclean}{\enesat\reclean}\;\Big[(\rhoeclean+\rhoesat-\reclean)^2-(\reclean)^2\Big]^2,
\label{clean-2}
\end{equation} 
and
\begin{equation}
\left\{
\begin{array}{rcl}
\enered(\tred)&\simeq&\frac{1}{2}+\Aredlog\ln\frac{\tred}{\tredhalfsat}\\
\frac{\dd\enered}{\dd\tred}&\simeq&\frac{\Aredlog}{\tred}
\end{array}
\right.\;\;
\begin{array}{c}
\mbox{(logarithmic or half-saturation }\\
\mbox{ regime $\enered\simeq\frac{1}{2}$),}
\end{array}
\label{sat-1}
\end{equation} 
\mbox{}\\
with
\begin{equation}
\Aredlog\;=\;\frac{(\,\omegabindingclean+\omegabindingsat\,)\;\tredhalfsat}{32\,\enesat\,\reclean}\;\Big[\,(\rhoeclean+\rhoesat-\reclean-\resat)^2-(\reclean+\resat)^2\,\Big]^2\,,
\label{sat-2}
\end{equation} 
\mbox{}\\
and \tredhalfsat\ defined by $\enered(\tredhalfsat)=1/2$ in \eq{diferencialred} (no truly explicit expression for \tredhalfsat\ seems to be feasible). In obtaining the above \eqsto{clean-1}{sat-2}, we have  assumed that $\enered(0)=0$ and that $\rhoe<\re$ at   $\tred=0$ or $\tredhalfsat$.

Unfortunately, to our knowledge no measurements exist of the time-evolution of the filtering efficiency of channels with nanostructured walls with a precision valid for a quantitative comparison with the corresponding results of our equations. We thus propose (see below) that such measurements should be made to further clarify the mechanism behind the enhanced impurity trapping capability of the chanels with nanostructured inner walls.

\mbox{}

\section{Conclusions and proposals for future work}

This Letter proposes a model for the main  generic features of the channels with nanostructured inner walls with respect to trapping and accumulation of impurities carried by fluids. This includes, \eg, their capability to clean the fluid from impurities of size well smaller than the channels' nominal radius, with comparatively small resistance to flow (much smaller than in a conventional channels of radius as small as the impurities). The model attributes the enhanced filtration capability to the long-range attraction exerted by the exposed charges in the nanostructured walls, and also to their binding capability once the impurities actually collide with them. Both features were quantitatively accounted for by means of a phenomenological  ``effective-charge density'' of the nanostructured wall. The model  also predicts an specific time-evolution of the trapped impurity concentration and of the filtering capability, which was  solved numerically, and also  analytically in the so-called clean and half-saturation approximations. 

We believe that our equations could make possible some  valuable future work, of which two specific matters seem to us more compelling: First, it would be interesting to check at the quantitative level the agreement with experiments of the  time-evolutions predicted above. For that, we propose to perform time-dependent measurements made in controlled flow setups. We have chosen in our \eq{diferencialred} flow constraints which seem appropriate for this purpose; still, the model can be tested with different setups by just numerically integrating  \eq{diferencial} under the corresponding experimental constraints. One of the prime purposes of this Letter is in fact to suggest such measurements.

A second interesting future work (already in progress in our research group) is  the design of optimal geometries for the combinations of channels  forming  filters. Our model opens this possibility because of the explicit use of  the Poiseuille relations, that  allow to calculate the resistance to flow of complex associations of those channels, in series and/or parallel. The effective diffusivity and tortuosity of the pathways network are also accounted for by these equivalent-circuit analyses.

\mbox{}\\
\mbox{}\\{\large\bf Acknowledgements.-- }\nonfrenchspacing This work has been  supported by the MICINN project FIS2010-19807 and by the Xunta de Galicia 2010/XA043 and 10TMT206012PR projects. All projects are co-funded by ERDF from the European Union. 

\newpage

\end{document}